\documentclass[aps,pra,showpacs,amsmath,amssymb,twocolumn,floatfix,floats]{revtex4}

\usepackage{graphicx}

\begin{document}

\title{Emission from dielectric cavities in terms of invariant sets of the chaotic ray
  dynamics  \\ ({ \small Phys. Rev. A 79, 013830 (2009)})}

\author{Eduardo G. Altmann}
\email{ega@northwestern.edu}
\affiliation{Northwestern Institute on Complex Systems, Northwestern University, Evanston,
  60628 IL, USA}


\begin{abstract}
The chaotic ray dynamics inside dielectric cavities is described by the
properties of an invariant chaotic saddle.  
The localization of the far-field emission in specific directions, recently observed in
different experiments and wave simulations, is found to be a
consequence of the filamentary pattern of the saddle's unstable manifold.
For cavities with mixed phase space, the chaotic saddle is divided in hyperbolic and
nonhyperbolic components, related, respectively, to the intermediate exponential
$(t<t_c)$ and the asymptotic power-law $(t>t_c)$ decay of the energy inside the cavity. 
The alignment of the manifolds of the two components of the saddle explains why even if
the energy concentration inside the cavity dramatically changes from~$t<t_c$ to~$t>t_c$,
the far-field emission changes only slightly. Simulations in the annular billiard confirm
and illustrate the main results. 
\end{abstract}


%


\maketitle

\section{Introduction}
Motivated by technological applications, the emission from dielectric microcavities of
different shapes has been the focus of detailed experimental
investigations~\cite{gmachl,lebental,swkim,tanaka,shim08} and wave
simulations~\cite{noeckel,lee,schwefel,leeR,shinohara,shinohara07,shim07,shinohara08,wiersig08}. 
Comparisons between the measurements and the predictions of the (chaotic) ray dynamics
reveal an overall good agreement, including detailed properties of the far-field
emission~\cite{schwefel,leeR,shinohara,wiersig08}.  
These recent results renew the interest on the classical (ray) dynamics in open chaotic
systems {\em per se}, i.e., not only as the short-wavelength limit of the quantum (wave)
description~\cite{casati,lu,keating,wiersig.weyl,nonnenmacher,keating.rmt,shepelyansky}.  

The ray dynamics inside dielectric cavities is determined by the laws of geometric optics: 
 rays travel in straight lines between collisions at the boundary of the cavity, where
 they generically split in reflected and transmitted (refracted) rays with intensities
 given by Fresnel's law.  
Far field emissions peaked in specific directions have been surprisingly observed even in
cavities where the reflected rays have (uniformly) chaotic
dynamics~\cite{gmachl,lebental,swkim,tanaka,noeckel,schwefel,shim07,wiersig08,wiersig06}. 
Directionality in the far field and good confinement (high $Q$ modes) are requirements for
applications of microcavities as lasing systems~\cite{gmachl}.  
The following two recent results have proved to be  crucial for a ray description of the
far field emission: 
(i) Lee {\em et al.}~\cite{lee} introduced the {\em survival probability distribution} of
the  intensities of rays inside the cavity.
In strongly chaotic systems, this distribution decays exponentially in time and numerical
evidence was presented for a steady phase-space dependence of the probability
distribution, independent of initial conditions~(see Ref.~\cite{ryu} for a detailed
description, and also Refs.~\cite{lee,leeR,shinohara,shinohara07,nonnenmacher}).  
(ii) {\em Schwefel et al.}~\cite{schwefel} explained the observation of peaked far-field
intensities by relating the regions of high emission to the unstable manifold of periodic
orbits close to the boundary of the region of total internal reflection.  
This has been further investigated in Refs.~\cite{shim08,leeR,shinohara,shim07,wiersig08}.

In this paper, the ray dynamics in dielectric cavities is described using the ergodic
theory of transient chaos~\cite{ott,gaspard,dorfman,TG}.  
After properly taking into account the {\em partial} leak characteristic of dielectric
cavities, the long-time properties of the chaotic dynamics are shown to be 
governed by an invariant set of the classical dynamics, the so called {\em chaotic saddle}
(CS), composed by all trajectories that never leave the cavity in both forward and
backward times. The results from the theory of transient chaos are then applied, what
gives the correct theoretical framework for the description of chaotic optical
cavities. In particular, a more general interpretation of the 
results (i) and (ii) mentioned above becomes apparent:   
(i) the steady survival probability distribution introduced by Lee {\em et al.}
in~\cite{lee} is equivalent to the conditionally invariant
density~$\rho_c$~\cite{PY,T,DY}; and (ii) the emission pattern is governed by the unstable
manifold of the CS (not only of a single periodic orbit) along which the c~measure
concentrates~\cite{T}.  
The importance of the CS and its manifolds in quantum open systems has been recently
recognized to explain the distribution of resonances~\cite{casati,keating,nonnenmacher} and as
the origin of a fractal Weyl's law~\cite{lu,wiersig.weyl,shepelyansky}.  
Here, instead, I focus on the importance of the CS for the {\em classical} ray dynamics
inside dielectric cavities, which are partially open systems. I find that the main
physical observables (decay rate~$\gamma$, emission pattern) can be obtained from 
properties of the CS.
Furthermore, I argue how to extend these results to the case of generic cavities, where
regions of regular and chaotic motion coexist in a {\em mixed phase space}. 
In particular, I show how  a division of the CS in hyperbolic and nonhyperbolic
components~\cite{JTZ,letter,longo} explains why even if the energy concentration inside
the cavity changes in time, the far field emission retains its main properties. 

The paper is divided as follows. In Sec.~\ref{sec.classicalrays} the classical ray
dynamics and the standard description in terms of the chaotic saddle is presented. 
The case of systems with mixed phase space is considered in Sec.~\ref{sec.nonH}.  
Sec.~\ref{sec.numerics} presents numerical simulations on the annular billiard.
Finally, the main conclusions are summarized in Sec.~\ref{sec.conclusion}

\section{Chaotic ray dynamics in dielectric cavities}\label{sec.classicalrays}

\subsection{Classical ray dynamics}

Rays inside a dielectric cavity travel in straight lines between successive collisions at
the cavity's boundary, where 
the ray generically splits in a reflected  and a transmitted (or refracted) ray. 
The direction of propagation of the rays are determined by the angle with respect to the
boundary's normal vector at the collision point.  
The reflected angle~$\theta_R \equiv \theta$ is equal to the incident angle~$\theta_I$,
while the transmitted angle~$\theta_T$ is given by Snell's law as $\sin \theta_T = n \sin
\theta_I$, where $n$ is the ratio between the (constant) refractive indices inside and
outside the cavity.  
The intensities of the rays after collision are given by Fresnel's law and total internal
reflection occurs for~$p\equiv \sin \theta_I > \sin \theta_c = 1/n \equiv p_c$.  
These are the well established laws of geometric optics. 

Assuming the validity of geometric optics, the dynamics of a ray is defined exclusively by
its initial condition and the geometry of the cavity's boundary (parametrized by~$s$).  
For simplicity, let us consider the case of two dimensional cavities or billiards (the
main results below remain valid for the three-dimensional case).  
The boundary's geometry defines a function~$M$ that maps one collision to the next~$M:
(s_t,p_t) \mapsto (s_{t+1},p_{t+1})$. The map $M$ preserves the area~$d\mu = ds dp = ds d\sin\theta$,
which establishes the analogy to Hamiltonian systems.  
Below, the dynamics of maps~$M$ that have at least one chaotic component are considered.
The discrete time $t$ can be related to the actual time using the mean time between
bounces $\pi A/S c$, where $A$ is the area of the billiard, $S$ is the perimeter, and $c$
is the speed of the ray. This is an approximation for individual rays~\cite{montessagne}.  

The above description determines the dynamics in closed billiards. 
In order to introduce the escape through the transmitted rays (according to Snell's and
Fresnel's laws), we consider that each ray has an intensity~$i$, with~$i_{t=0}=1$.  
After each collision the intensity of the reflected ray~$i_{t+1}$ depends on~$i_t$, the
angle~$\theta_I$ and on the  polarization of the incident ray. For transverse magnetic
(TM) and transverse electric (TE) polarizations, the reflection coefficient~$R$ is
given by the square of Fresnel's coefficients
\begin{equation}\label{eq.erre}
\begin{array}{ll}
R_{TM}(\theta) =  \left[\frac{\sin(\theta_T-\theta_I)}{\sin(\theta_T+\theta_I)}\right]^2, \\
R_{TE}(\theta) =  \left[\frac{\tan(\theta_T-\theta_I)}{\tan(\theta_T+\theta_I)}\right]^2,
\end{array}
\end{equation}
for~$|\sin(\theta_I)|< 1/n=p_c$, and $R=1$ otherwise (total internal reflection).  
The transmitted rays have angle~$\theta_T$ and intensity~$T=1-R$. The region of the phase
space~$-p_c<p<p_c$, where~$T > 0$, will be denoted as {\em leak region}~$I$. 
The reinjection of transmitted rays into the cavity is neglected (it may occur in concave
billiards).

In summary, the full ray dynamics is given by $(s_t,p_t,i_t) \mapsto
(s_{t+1},p_{t+1},i_{t+1})$, where $M: (s_t,p_t) \mapsto (s_{t+1},p_{t+1})$ is an area
preserving map defined by the geometry of the billiard and the intensity $i_{t+1}=R(p)i_t$
decreases in time according to Fresnel's law~(\ref{eq.erre}).   
The energy in one region~$\Omega$ of the phase space at time~$t$ is given by the
intensities~$i_t$ and density~$\rho(s,p,t)$ of rays inside it:  
\begin{equation}\label{eq.energy}
E((s,p)\in\Omega,t)=\int_\Omega \int_\Omega  i(s,p,t) \rho(s,p,t) ds dp.
\end{equation}
The direction, position, and intensity of the rays emitted from the cavity can be computed
by Fresnel's and Snell's law from~$E(s,p,t)$ inside~$I$. 

\subsection{Estimations based on the closed system}

Let us first consider the case of billiards where the dynamics of the closed map~$M$ is
ergodic and 
strongly chaotic (e.g., uniformly hyperbolic)~\cite{ott}, leaving the generic case of
systems with mixed phase space for Secs.~\ref{sec.nonH} and~\ref{sec.numerics}.   
For strongly chaotic systems, after a short transient time~$t^*$ the fraction of rays that
never entered~$I$ decays exponentially~\cite{ott,TG}.  
The intensity~$i$ decreases with successive bounces inside~$I$, and rays in~$I$ return
typically exponentially fast to it~\cite{letter}. 
Therefore, the total energy~$E(t)$ inside the cavity  [i.e., considering $\Omega$
in Eq.~(\ref{eq.energy}) to be the full phase space] also decays 
exponentially~\cite{shinohara07,ryu}. This exponential decay is generically written as
\begin{equation}\label{eq.exp}
E(t) \sim (1-r)^t = \exp[\ln(1-r) t],
\end{equation}
where $"\sim"$ indicates that both sides of the relation approach a constant for long
times, and the constant leakage rate~$r$ corresponds to the transmitted energy per unit of
time~\cite{ott}. The escape rate~$\gamma$ of Eq.~(\ref{eq.exp}) is defined as
\begin{equation}\label{eq.gamma}
\gamma \equiv - \ln(1-r) \;\; [\approx r \text{  for small  } r].
\end{equation}
The ergodicity assumption for the closed system means that its phase space cannot be
divided in two dynamically disjoint
regions~$A$ and $B$ with $\mu(A)>0$ and~$\mu(B)>0$. 
Any initial density of rays~$\rho_0(s,p)\equiv\rho(s,p,t=0)$ converges (exponentially
fast for strongly chaotic systems) to the natural density~$\rho_\mu$ (constant in
the phase space area $dp ds$).
When the leak~$I$ is small, a popular simplifying assumption is to consider the rays
inside the {\em open} billiard at a given long time~$t$ to be distributed according to~$\rho_\mu$,
i.e., according to the natural measure~$\mu$ of the {\em closed} billiard. Under this
assumption, and taking into account that the transmission at time~$t$ occurs from
inside~$I$ according to~$T=1-R$, an approximation~$r^*$ for the leakage rate~$r$
in Eq.~(\ref{eq.gamma}) can be calculated as   
\begin{equation}\label{eq.random}
 r^* \approx \int_I T(\theta) d\mu =  \int_0^1 \int_{-\theta_c}^{+\theta_c}
 [1-R(\theta)] \cos \theta d\theta ds. 
\end{equation}
The leakage rate~$r$ was called the {\em degree of leakage} by Ryu {\em et al.} in Ref.~\cite{ryu}, where
analytical expressions for approximation~(\ref{eq.random}), using~$R_{TM}$ and~$R_{TE}$ given
by Eq.~(\ref{eq.erre}), were calculated.  
Ryu {\em et al.} show the interesting dependency~$r\sim 1/n^2$ that was verified numerically.  
Similarly, the ray dynamics described above has been successfully applied in cavities with
different
shapes~\cite{gmachl,lebental,swkim,tanaka,noeckel,lee,schwefel,leeR,shinohara,shinohara07,shinohara08,wiersig08}. 
It is interesting to compare these applications in optics to previous investigations
involving other systems with leaks~\cite{PY,alt,lai,paar,schneider,nagler.billiards,letter}.  
The main difference is the {\em partial} leak through Fresnel's law in the optical systems
(also present in acoustics~\cite{montessagne}), in opposition to a complete escape assumed
in the previous cases. 
However, as we will see below, once the intensity of the rays is properly taken into
account, a complete correspondence can be established.  
For instance, relation~(\ref{eq.random}) is a standard estimation~\cite{ott,dorfman,TG}. 
Dependence on the position of~$I$ has been reported~\cite{paar,schneider} and, as
expected, Eq.~(\ref{eq.random}) is strictly valid only in the limit of~$r \rightarrow 0$,
which is of little practical interest for optical systems since it corresponds
to~$n\rightarrow\infty$.  
However, it is important to note that approximations based on {\em closed} system's
properties, such as the one leading to Eq.~(\ref{eq.random}),
are often the only available ones and lead to successful predictions~\cite{lebental}. 
In the next section, the analogy to leaked systems is deepened and
fundamental results of the ergodic theory of transient chaos (and chaotic scattering) are
used to
obtain a description of ray dynamics that fully incorporates the openness of optical cavities.

\subsection{Description in terms of invariant sets of the open system}\label{ssec.invariantsets}

Transient chaotic motion is typical in systems with leaks and in naturally open systems showing
chaotic scattering~\cite{gaspard,dorfman,ott,TG}.  
The escape of trajectories that remain a long time inside the system is governed by an
invariant, nonattracting, chaotic set~\cite{gaspard,dorfman,ott,TG}. 
This set is composed by the trajectories that never leave the system, neither in forward
nor in backward iterations of the map.  
The stable (unstable) manifold of this set is defined by all points that lead to this set
in forward (backward) time.  
The term {\em chaotic repeller} is sometimes used to denote this
set~\cite{gaspard,dorfman,wiersig.weyl,nonnenmacher,wiersig08}.  
Because billiards have a two dimensional phase space and are time reversible, having therefore stable and unstable
manifolds, the term {\em chaotic saddle} (CS) is more appropriate~\cite{T,TG}.  
For strongly chaotic ergodic systems, the CS has zero Lebesgue measure $\mu(CS)=0$
(vanishing area of the phase space), and the support of the CS is a fractal set. The
stable and unstable manifolds cross orthogonally (angle bounded from zero) and are also of
zero Lebesgue measure.  
Trajectories that survive for a long time inside the system necessarily have initial
conditions close to the stable manifold of the CS, approach closely the CS, and leave the
system through the unstable manifold of the CS. 
A well defined escape rate~$\gamma$ exists which is independent of the density of initial
conditions~$\rho_0$, provided the support of~$\rho_0$ intersects the stable manifold of the CS.   
Relations between~$\gamma$ and the properties of the CS  (fractal dimension along the
manifolds, Lyapunov exponent) have previously been derived~\cite{ott,TG,dorfman}. 

Let us see now how these results can be adapted to the case of optical cavities, where the
leakage is only partial inside~$I$ and the dynamics involves not only the map~$M$ but also
the decay of the intensity~$i$.  
A natural definition of the CS is obtained replacing the condition of never escaping
trajectories mentioned above by the condition that $i=1$ for all times: 
\begin{equation}\label{eq.cs1}
(p_{CS},s_{CS}) \in \text{CS} \Leftrightarrow i(p_{CS},s_{CS},t\rightarrow\pm\infty)=1.
\end{equation}
In other words, the CS defined Eq.~(\ref{eq.cs1}) is the same obtained
considering a full leak and the standard definition of the saddle, i.e., replacing the
partial leak in Eq.~(\ref{eq.erre}) by a Heaviside step function. It follows that: CS$~\cap~I
= \emptyset$ because escape takes place if $(p,s) \in I$.
A definition similar to~(\ref{eq.cs1}) is not appropriate for the manifolds of the CS.  
For instance, initial condition inside~$I$ that converge to the CS
for~$t\rightarrow\infty$ and still have~$i_{t\rightarrow\infty}\neq 0$ clearly deserve to
belong to the saddle's stable manifold (similar argument for~$t\rightarrow -\infty$ holds
for the unstable manifold).  
The stable (unstable) manifold of the CS is therefore defined by {\em all} points $(s,p)
\rightarrow$ CS for~$t\rightarrow +\infty$ ($-\infty$), but attached to these points there
is a manifold intensity~$i$ given by $i_{t\rightarrow\infty}$
($i_{t\rightarrow-\infty}$). 
This suggests an alternative definition for the CS itself:
\begin{equation}\label{eq.cs2}
(p_{CS},s_{CS}) \in \text{CS} \Leftrightarrow i(p_{CS},s_{CS},t\rightarrow\pm\infty)>0.
\end{equation}
The CS defined using Eq.~(\ref{eq.cs2}) contains the CS defined using Eq.~(\ref{eq.cs1}):
CS-(\ref{eq.cs1}) $\subset$ CS-(\ref{eq.cs2}).  
Moreover, all points in CS-(\ref{eq.cs2}) but not in CS-(\ref{eq.cs1}), collide only
a finite number of times inside~$I$ and necessarily belong to the intersection of the
stable and unstable manifolds of CS-(\ref{eq.cs1}). 
Therefore, for long times, all rays with nonvanishing intensities inside the cavity will
be governed by the CS~(\ref{eq.cs1}) and its manifolds. 
This justifies the choice of relation~(\ref{eq.cs1}) and shows that this CS also governs
the energy decay from the billiard.

With the above definitions of the invariant sets, let us characterize the escape from the
cavity.  
The proper measure to describe this decay is the conditionally invariant measure (c
measure) $d\mu_c$~\cite{PY,T,DY}.  
Intuitively, the mathematically well defined c measure is obtained multiplying the
survival density by a factor proportional to~$\exp(\gamma t)$ that compensates the decay
of the Lebesgue measure~\cite{letter}.  
The conditionally invariant density~$\rho_c$ concentrates along the unstable manifold of
the CS~\cite{T} and is the only attractor for typical initial
densities~$\rho_0$~\cite{PY,DY}.  An important property of the c measure is that it
converges to the natural measure for small leak regions $\mu(I)\rightarrow 0$~\cite{PY,DY}, which justifies
the approximation used to obtain Eq.~(\ref{eq.random}). However, typical dielectric cavities
have~$n<10$ and the approximation of small leak is violated. This means that the dynamical
properties derived from the geometry of the (closed) billiard are not a good approximation
to the dynamics of the optical (open) billiard. In analogy to the calculation
in Eq.~(\ref{eq.random}), but this time using the precise distribution inside the cavity given by~$\rho_c$,
the leakage rate~$r$ in Eq.~(\ref{eq.exp}) is given by the c measure of the leak~$I$~\cite{PY,paar2,letter}  
\begin{equation}\label{eq.cr}
 r =  \int_I T(\theta) d\mu_c = \int_0^1 \int_{-\theta_c}^{+\theta_c} [1-R(\theta)]
 \rho_c(\theta,s) \cos \theta d\theta ds. 
\end{equation}
The escape rate~$\gamma$ is obtained from Eq.~(\ref{eq.gamma}). For small~$I$ (large
$n$)~$\rho_c$ approaches~$\rho_\mu$ and $r$ in~(\ref{eq.cr}) approaches $r^*$, obtained 
in~(\ref{eq.random}). Typically $r^*$ overestimates~$r$~\cite{longo}.  

\begin{figure*}[tb]
\includegraphics[width=0.95\columnwidth]{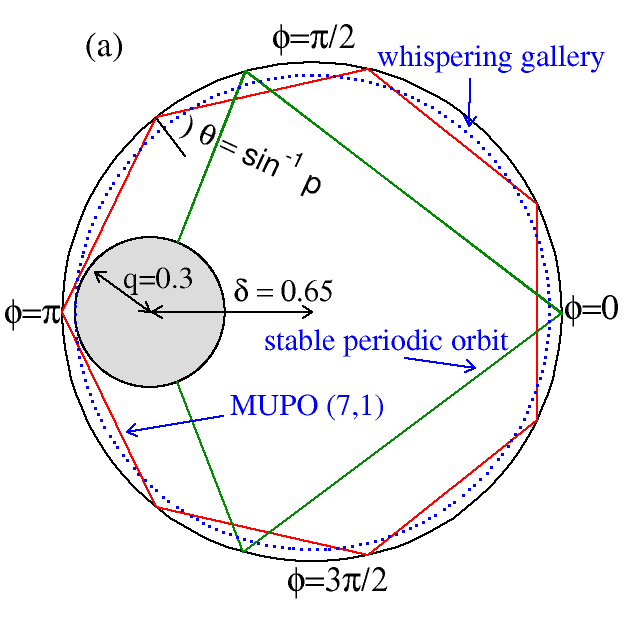}\hspace{0.1\columnwidth}\includegraphics[width=0.95\columnwidth]{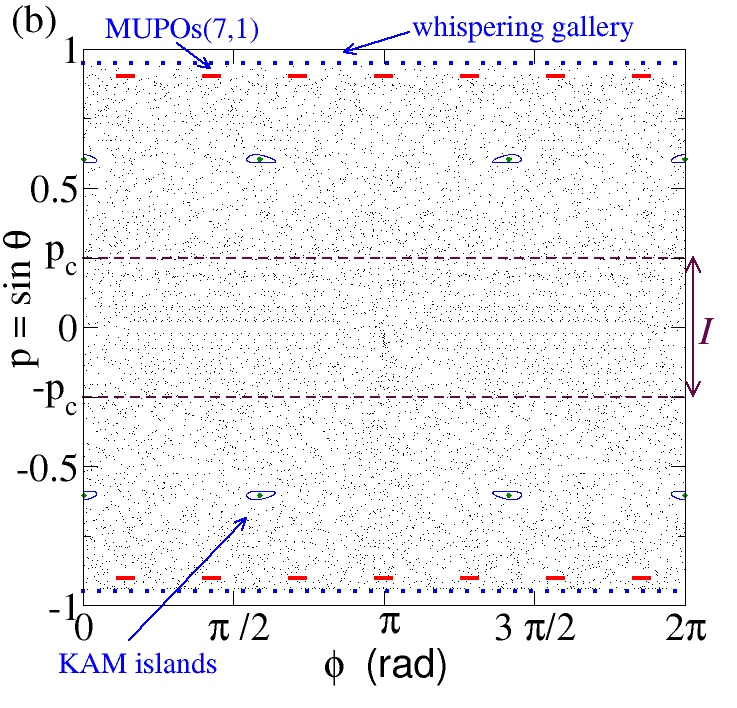}  
\caption{(Color online) (a) Annular billiard defined by two eccentric circles, with
  control parameters $q=0.65$ and~$\delta=0.3$.
Lines indicate one stable periodic orbit (at $\phi =0$) and one MUPO~(7,1) (polygonal
form)~\cite{mupos}.  
Trajectories that do not cross the dotted circumference of radius~$q+\delta$ belong to the
whispering gallery.      
(b) Phase space representation of the closed billiard shown in (a), obtained using a
Poincar\'e surface of section at the outer boundary (parametrized by~$s=\phi\in[0,2\pi]$).  
The stable periodic orbit in (a) is at the center of a chain of KAM islands. Black dots
mark $8000$ iterations of a single chaotic trajectory. The horizontal dashed lines
denote~$p_c= \pm 1/n$ for $n=4$, as used in the following figures.} 
\label{fig.1}
\end{figure*}

It is remarkable that many of the results of the theory of transient chaos presented above have been recently rediscovered
in the analysis of optical cavities. 
For instance, the steady state of the survival probability distribution, introduced by Lee
{\em et al.}~\cite{lee} and mentioned as point (i) in the Introduction, is
equivalent to the conditionally invariant density~$\rho_c$ introduced by Pianigiani and
Yorke thirty years ago~\cite{PY} (see Refs.~\cite{DY,T} for a recent review).  
The independence of initial ensembles reported in Ref.~\cite{lee} is related to the
existence of the invariant CS or, equivalently, to the fact that~$\rho_c$ is the only
attractor for typical~$\rho_0$'s.  
The density~$\rho_c$ concentrates along the unstable manifold of the CS~\cite{T}, which
presents the characteristic filamentary pattern inside~$I$.  
This explains the peaked distribution of the transmitted rays responsible for the emission
from the cavity. 
When a short time periodic orbit exists close to the critical line~$p=p_c$, the unstable
manifold of this orbit (which also belongs to the CS) will be parallel to the manifold of
the remaining part of the CS because both manifolds do not intersect.  
This corresponds to the observation by Schwefel {\em et al.}~\cite{schwefel} described as
point (ii) in the Introduction.

\section{Systems with mixed phase space}\label{sec.nonH}

The results described so far can be rigorously applied only for a limited class of
strongly chaotic systems.  
It can be argued that it is a technical problem to extend these demonstrations to a larger
class of nonuniformly hyperbolic systems, where no deviation of the exponential decay have
been numerically detected. 
However, cavities with generic boundaries typically have a mixed phase space: coexisting
with regions of chaotic motion there are regions of regular motion, e.g.,
Kolmogorov-Arnold-Moser (KAM) islands. 
Around these regions there is a {\em sticky region}  whereto chaotic trajectories get
partially trapped, introducing a power-law like decay of the survival probability.  
Rigorously, the results of the theory of transient chaos mentioned above either do not
apply or become trivial. 
This type of  cavities have been considered in Refs.~\cite{noeckel,schwefel,swkim} and
without further justifications it is not clear how the previous results can be extended to
this case.  
 
In this section, I show how, despite the mathematical difficulties, in practice the
formalism of the CS and its manifolds do apply to billiards presenting a divided
phase space.  
The basic observation is that typically a well defined exponential decay of the survival
probability exists for intermediate times, where in practice the previous results can be
applied~\cite{JTZ,letter}. 
Below, billiards containing a large chaotic component and no KAM islands in the
border of~$I$ are considered.
In this case, similar to the decay of trajectories, the decay of energy~(\ref{eq.energy})
from the chaotic component shows a transition from exponential to asymptotic
power-law~\cite{alt,lai,ryu,nagler.billiards,letter,sharply} 
\begin{equation}\label{eq.powerlaw}
E(t) = \left\{ \begin{array}{ll} A e^{-\gamma t} & \text{ for } t > t^*, \\
      A e^{-\gamma t} + B t^{-\alpha} & \text{ for } t > t_\alpha,
\end{array}
\right.
\end{equation}
where~$t^*$ is proportional to the inverse of the negative Lyapunov exponent of the
saddle, $t_\alpha>t^*$ is the shortest time rays in the sticky region need to reach $I$, and
$A e^{-\gamma t_\alpha} \gg B t_\alpha^{-\alpha}$. 
The physically relevant time is the crossover time~$t_c > t_\alpha$ between the exponential and power-law decays
in Eq.~(\ref{eq.powerlaw}). It is defined as~$A e^{-\gamma t_c} = B t_c^{-\alpha} = E(t_c)/2$,
i.e., for~$t>t_c$ the power-law decay dominates. 
In Ref.~\cite{letter}  it was shown that~$t_c\sim 1/\gamma$ and suggested that exponential
and power-law regimes in Eq.~(\ref{eq.powerlaw}) can be related, respectively, to hyperbolic
and nonhyperbolic components of the CS, as first suggested in Ref.~\cite{JTZ}.  
The nonhyperbolic component consists of the border of KAM islands (and by other marginal
stable orbits) while the hyperbolic part is away from the sticky regions and resembles the
CS described in Sec.~\ref{ssec.invariantsets}. 
Initial conditions uniformly distributed {\em touching} the KAM islands reduces the
exponent~$\alpha$ in Eq.~(\ref{eq.powerlaw}) by~$1$ (see Ref.~\cite{sharply} and references
therein). Considering this effect, and because~(\ref{eq.powerlaw}) 
is a survival probability distribution, typically~$0< \alpha \leq 1$.  
The exact (finite time) value depends on the properties of the nonhyperbolic region (KAM
island).

In terms of the energy inside the cavity, Eq.~(\ref{eq.powerlaw}) indicates that while
for~$t<t_c$ energy concentrates strongly in the hyperbolic component of the CS and its
manifolds, for~$t>t_c$ the energy concentrates in the nonhyperbolic component of the CS
close to the KAM islands.  
In principle, one could expect also a dramatic change in the emission pattern from $t<t_c$
to~$t>t_c$.  
However, as emphasized in Ref.~\cite{letter}, one important difference between the
hyperbolic and nonhyperbolic components of the CS is that their manifolds attract and
repel exponentially and sub-exponentially respectively.  
Therefore, rays slowly approach and slowly escape the nonhyperbolic component of the CS {\em
  through} the hyperbolic component.  In particular, when KAM islands are away from~$I$,
rays that approached KAM islands will come close to the 
stable/unstable manifolds of the {\em hyperbolic} component of the CS before escaping.
More precisely, the unstable manifold of the nonhyperbolic component of the CS aligns to
and follows 
closely the unstable manifold of the hyperbolic component. 
From this qualitative description we expect that, even if the energy concentrates for long times~$t>t_c$
on the nonhyperbolic component of the CS, the decay towards~$I$ will not dramatically
change from $t<t_c$ to $t>t_c$. 
This prediction is partially confirmed in the next Section, where numerical simulations of
a specific system are presented. 

\section{Numerical simulations}\label{sec.numerics}

\subsection{The annular billiard}

Numerical simulations of the ray dynamics are performed in the annular billiard, composed
by two circles (of radii~$q<1$ and~$R=1$) placed eccentrically (distance $\delta$ between
centers), as depicted in Fig.~\ref{fig.1}a.  
The closed phase space shown in Fig.~\ref{fig.1}b shows a large chaotic sea coexisting
with three regions of regular dynamics: the chain of KAM islands and two whispering
gallery regions close to the outer boundary.  
Whispering gallery regions are typical for billiards with concave
shape~\cite{noeckel,wiersig08,tanaka}.  
In the case of the annular billiard an infinite number of families of marginally unstable
periodic orbits (MUPOs) accumulate outside of the whispering gallery~\cite{mupos}, whereto
the chaotic trajectories stick with~$\alpha=1$ in Eq.~(\ref{eq.powerlaw})~\cite{sharply}.

\subsection{Time dependence}\label{ssec.time}

\begin{figure}
\includegraphics[width=1\columnwidth]{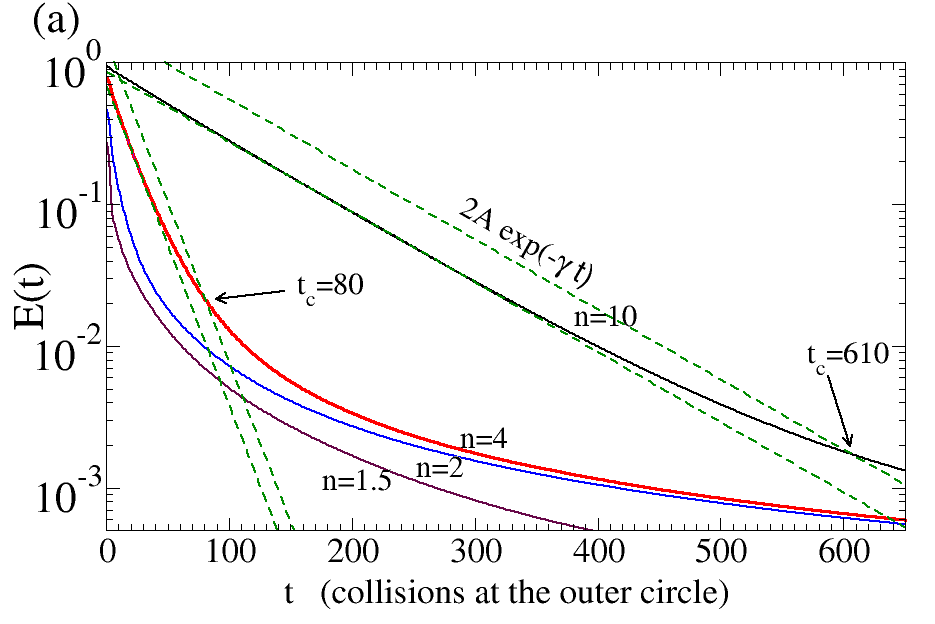}\\\vspace{0.2cm}
\includegraphics[width=1\columnwidth]{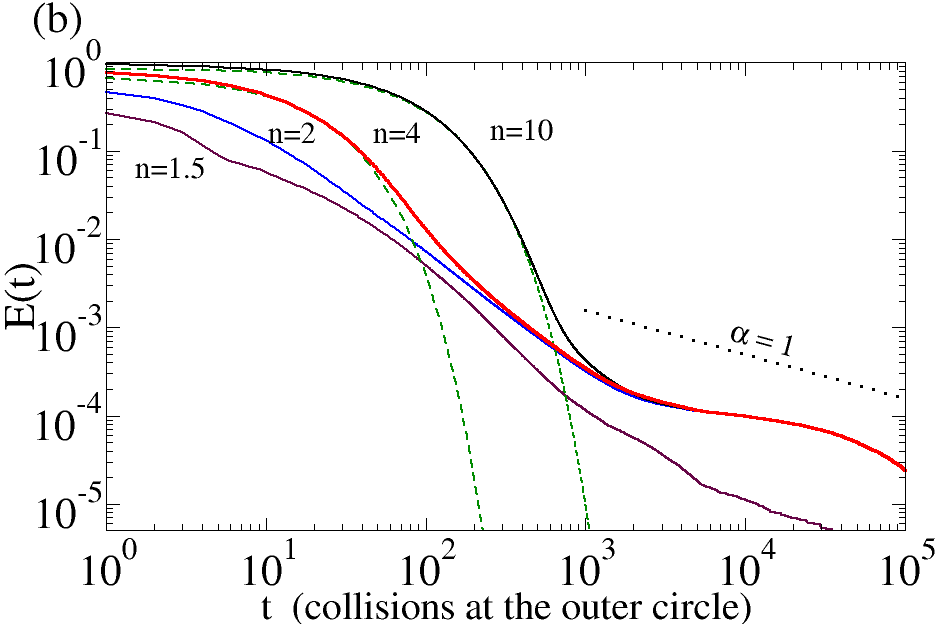}
\caption{(Color online) Fraction of initial energy inside the billiard shown in
  Fig.~\ref{fig.1} as a function of time. Different curves correspond to the refractive
  indices $n=\{10,4,2,1.5\}$ (from top to bottom). (a) Linear-log and (b) log-log scale. The
  dashed lines correspond to an exponential fitting for short times and $t_c$ indicates
  the transition time to a power-law. Rays were started uniformly distributed in the
  chaotic component.} 
\label{fig.2}
\end{figure}

\begin{figure}
\includegraphics[width=0.9\columnwidth]{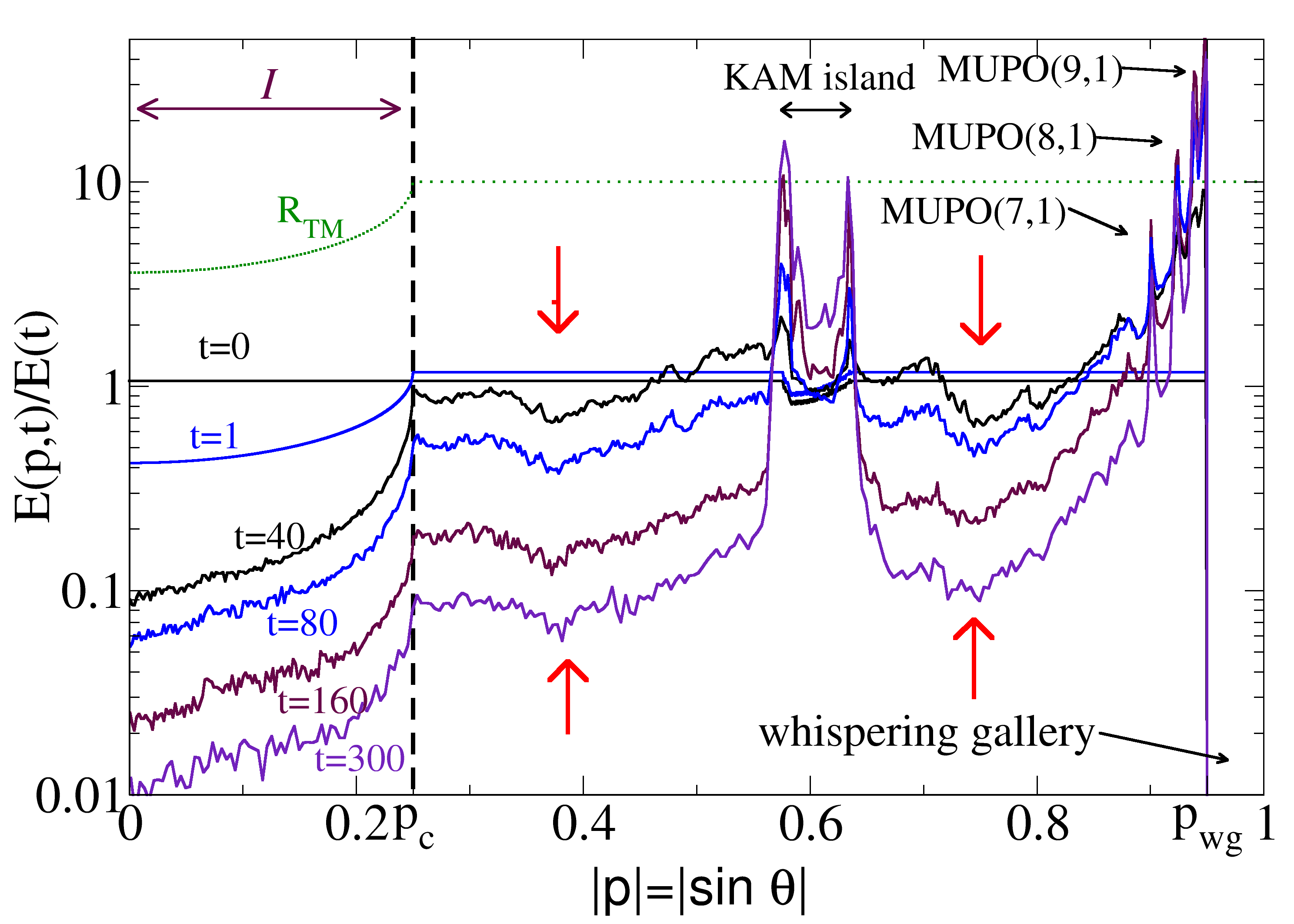}
\caption{(Color online) Energy along~$p$ at time~$t$ normalized by total energy at
  time~$t$, for $t=0,1,40,80,160,$ and~$300$ (top to bottom).  
Refractive index $n=4$ leads to~$p_c=0.25$ and $t_c=80$ (see Fig.~\ref{fig.2}).  
Rays were started uniformly distributed in the chaotic component ({\em outside} the KAM
island). Vertical arrows indicate minima of the distribution that systematically appear for
all times. The dotted line indicates the reflection coefficient~$R_{TM}$ in Eq.~(\ref{eq.erre}),
multiplied by $10$ for clarity.} 
\label{fig.3} 
\end{figure}

Consider now that the larger circle in the annular billiard~(Fig.~\ref{fig.1}a) has
refractive index~$n$, the outside space~$n=1$, and the boundary of the inner circle is a
perfect mirror.  
In this case, emission according to Eq.~(\ref{eq.erre}) is possible through the outer border
when~$|p| < |p_c| = 1/n$, i.e., the leak region~$I$ corresponds to the horizontal stripe
between $-p_c < p < p_c$ in the center of the phase space shown in Fig.~\ref{fig.1}(b).  
Typically,~$10^7$ TM polarized rays (similar results are expected for TE polarization) are started
 uniformly distributed according to the Lebesgue measure (area
of the phase space) inside the chaotic component.  
This means that no rays are started inside the KAM island or whispering gallery.  
The temporal decay of the total energy is presented in Fig.~\ref{fig.2}  for different
values of the refractive index~$n$.  
It confirms the existence of an intermediate exponential decay and an asymptotic power-law
decay, as described by Eq.~(\ref{eq.powerlaw}).  
Different exponents can be identified for~$n=2$ and~$4$: $1<\alpha\leq 2$ for intermediate
times, related to rays that approach the KAM island (started away from it), and~$0< \alpha
\leq 1$ related to rays started already close to the KAM islands.  
For~$n=2,4,$ and~$10$ the energy decay converges precisely to the same curve
for~$t\rightarrow\infty$, because 
the nonhyperbolic component of the CS is the same in all these cases. For $n=1.5$
this is not true because the KAM island chain is (partially) inside~$I$.
Hereafter the representative case~$n=4$ is chosen, for which the generic exponential and
power-law decays are clearly visible in Fig.~\ref{fig.1} and~$t_c=80$. 
 
Let us see now how the energy is distributed inside the billiard at a given time~$t$.  
Figure~\ref{fig.3} shows the distribution projected to the~$p$ axis for different times
$t$. 
 The total energy in all cases is normalized and the rays are started uniformly in the chaotic component. 
 For a fully chaotic system this distribution converges to the conditionally invariant
 density~$\rho_c$ (see Sec.~\ref{sec.nonH}) projected on the $p$-axis.  
 Instead, Fig.~\ref{fig.3} shows a concentration of the energy close to the KAM island and
 whispering gallery. 
This is a consequence of the power-law escape of rays in these sticky regions, in
opposition to the exponential escape for rays away from these regions.  
Apart from a rescale due to this decay, Fig.~\ref{fig.3} suggests that the
distribution away from islands also follows a similar pattern (see, e.g., the vertical arrows
in Fig.~\ref{fig.3}) for all times~$t<t_c$ and~$t>t_c$.  
This is in agreement with the interpretation given at the end of Sec.~\ref{sec.nonH} and
will be further investigated in Sec.~\ref{ssec.edist}.

\subsection{Chaotic saddle}\label{ssec.cs}

\begin{figure}[tb]
\includegraphics[width=0.9\columnwidth]{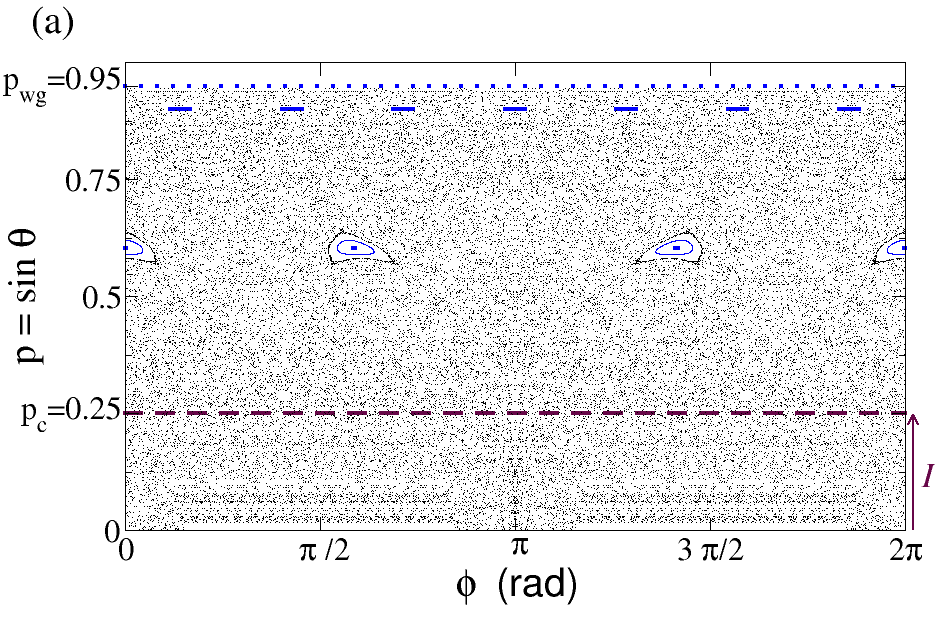}\\
\includegraphics[width=0.9\columnwidth]{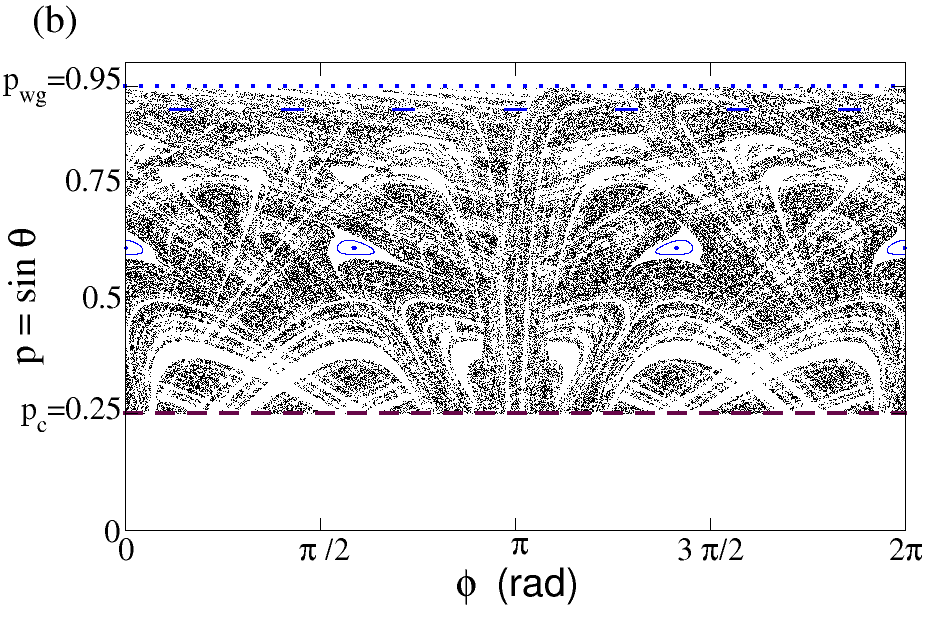}\\
\includegraphics[width=0.9\columnwidth]{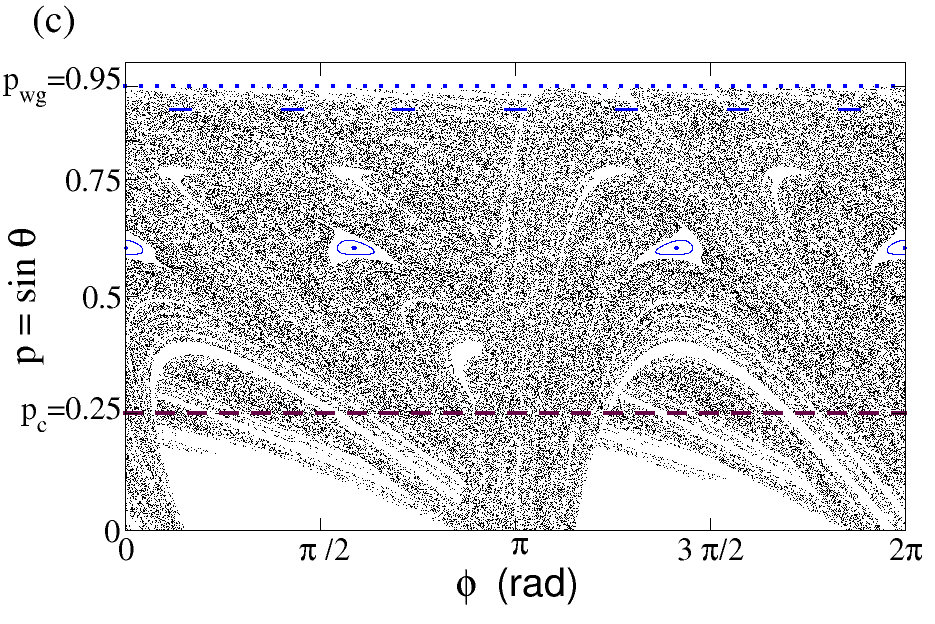}
\caption{(Color online) (a) Closed system's phase space [magnification of
  Fig.~\ref{fig.1}(b)]. (b) Hyperbolic component of the CS, and (c) its unstable manifold
  for $n=4$. The method of Ref.~\cite{TG} with $t^\dagger=30<t_c=80$ was employed, with rays
  started away from the regions of regular motion. Notice that the
  unstable manifold enters~$I$ in (c), a sign of rays leaving the cavity (only points
  with~$i>0.01$ are plotted). Due to the time reversible symmetry and the spatial
  symmetry of the annular billiard, the stable manifold of the CS can be obtained from the
  unstable one by~$(s,p)\mapsto(2\pi-s,p)$. } 
\label{fig.4}
\end{figure}

\begin{figure*}[tb]
\includegraphics[width=0.95\columnwidth]{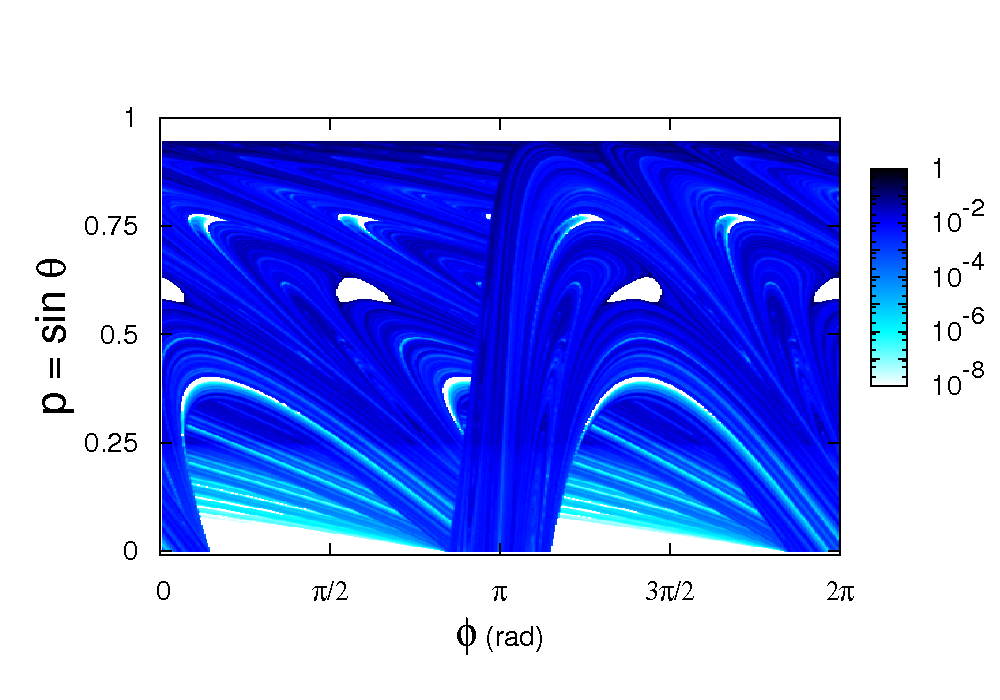}\includegraphics[width=0.95\columnwidth]{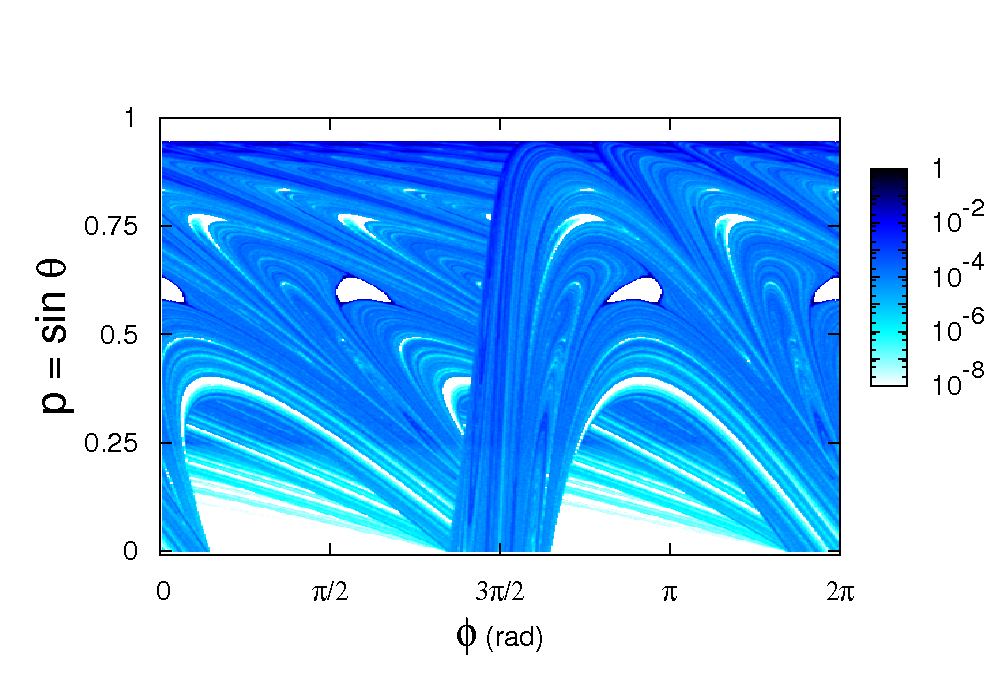}\\
\includegraphics[width=0.95\columnwidth]{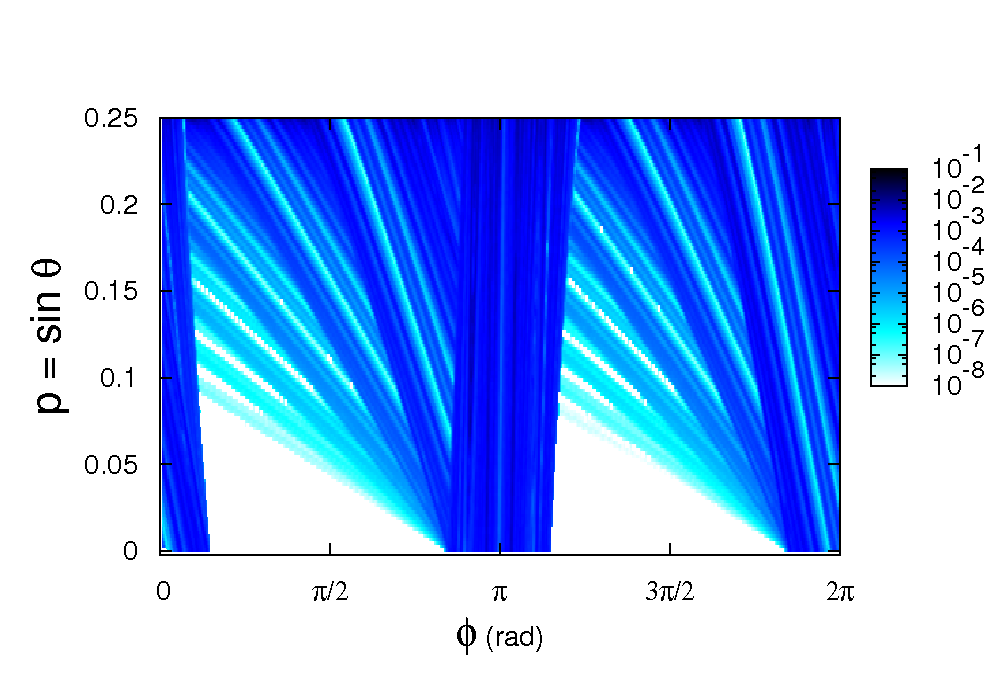}\includegraphics[width=0.95\columnwidth]{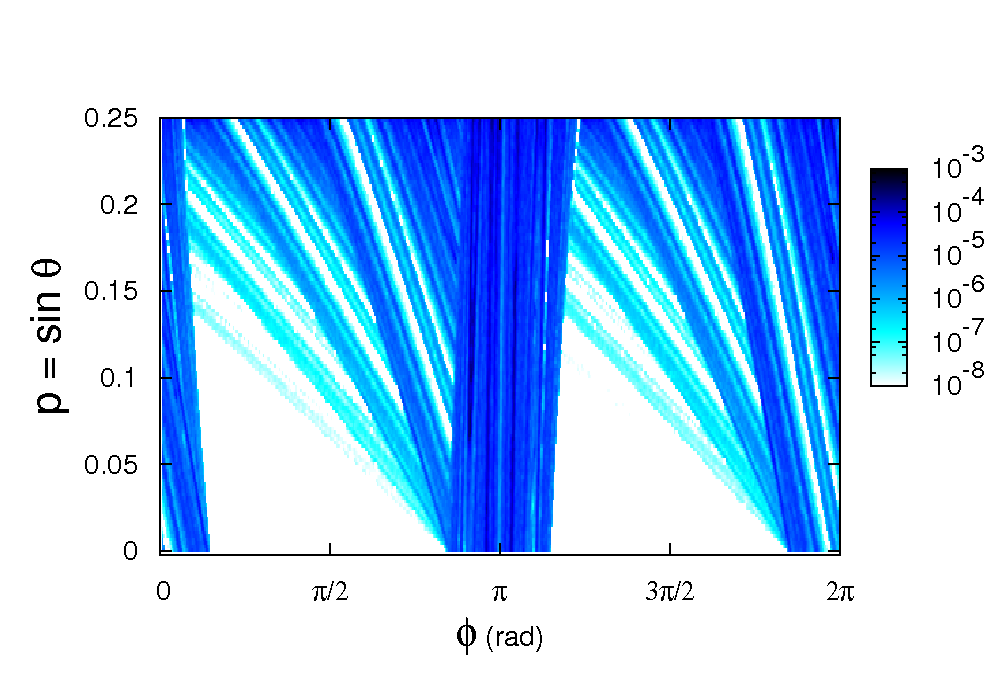}
\caption{(Color online) Energy density in the phase space of the annular billiard
  for $n=4$ ($p_c=0.25$) at two different times: Left $t=40=t_c/2$; Right
  $t=160=2t_c$. Bottom row shows magnifications of the top row in the leak region. Rays
  were started uniformly distributed in the chaotic region.} 
\label{fig.5}
\end{figure*}

\begin{figure}
\includegraphics[width=1\linewidth]{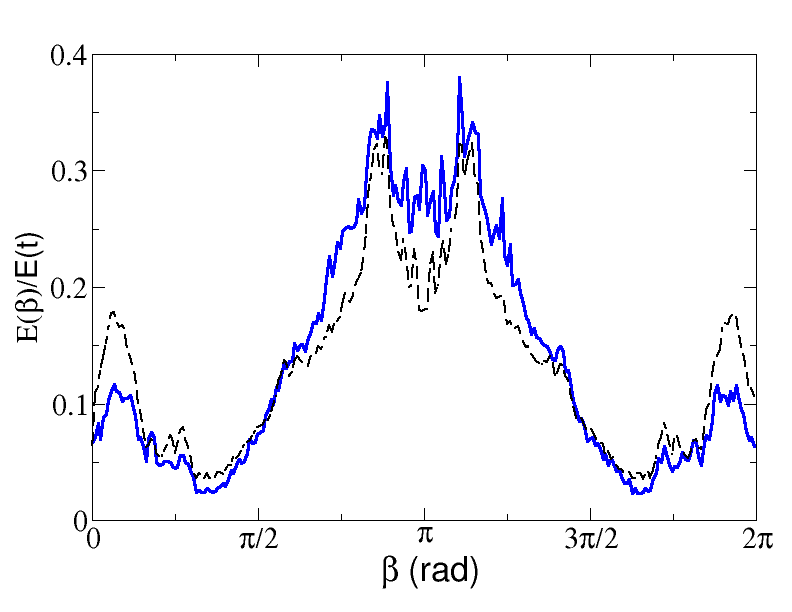}
\caption{(Color online) Far field intensity transmitted in the
  direction~$\beta=\phi-\theta_T$ measured from the center of the annular billiard at
  times~$t=t_c/2=40$ (dashed line) and $t=2 t_c =160$ (solid line).  Rays were started
  uniformly distributed in the chaotic region and~$n=4$.} 
\label{fig.6}
\end{figure}

In this section we will see how the division of the chaotic saddle in hyperbolic and
nonhyperbolic components proposed in Sec.~\ref{sec.nonH}  
apply for the annular billiard considered here.
The nonhyperbolic component of the CS is composed by the border of the KAM islands, the
border of the whispering gallery, and the families of MUPOs.  
Figure~\ref{fig.3} shows that the energy concentrates in this region for long times. 
Regarding the hyperbolic component, a simple and efficient method to obtain a
visualization of this zero measure set was proposed in Ref.~\cite{TG} (p. 201).  
It is based on the observation that (most) trajectories that survive until some long
time~$t^\dagger$  were close to the stable manifold of the CS at time $t=0$, were close to
the CS at time~$t=t^\dagger/2$, and were close to the unstable manifold at
time~$t=t^\dagger$.  
This method is expected to work for the hyperbolic component of the CS if~$t^*\ll
t^\dagger < t_c$ and if initial conditions are selected away from sticky regions, i.e.,
{\em non-uniformly} in the chaotic sea.  
The results achieved for the annular billiard are shown in Figs.~\ref{fig.4}(b) and \ref{fig.4}(c). 
Comparing to the closed system's phase space shown in Fig.~\ref{fig.4}(a), it is evident
that non-trivial structures were created by the openness of the system.
Furthermore, in opposite to the density~$\rho_\mu$ of the closed system [support shown in
Fig.~\ref{fig.4}(a)], the conditionally invariant density~$\rho_c$ of 
the open system [support shown in Fig.~\ref{fig.4}(c)] is not smooth.

\subsection{Energy distribution and emission}\label{ssec.edist}

Finally, the energy distribution inside the cavity and the far field emission are
investigated.  
Considering the time dependency obtained in Sec.~\ref{ssec.time}, where $t_c=80$ was found
for~$n=4$, two times are considered:  
$t=t_c/2=40$, for which the decay is exponential and the hyperbolic component of the CS
dominates, and $t=2 t_c = 160$, for which the decay is algebraic and the nonhyperbolic
component of the CS dominates. 
The relative (apart from the overall decay) phase-space distribution of energy at these
times are expected to be representative for 
all times in the exponential and power-law decays  (e.g., the normalized  distribution
for~$t\rightarrow\infty$ is expected to resemble the one at~$t=160$). 
The numerical results are shown in Fig.~\ref{fig.5}. 
The two figures in the upper row confirm that the energy shifts from the hyperbolic
component of the CS and its unstable manifold, to the nonhyperbolic component of the
saddle.  
The two figures in the bottom row are magnifications of the upper figures in the leak
region, which is the relevant region for emission purposes.  
The differences between $t=40$ and~$t=160$ are much less dramatic in this case. Comparing
to Fig.~\ref{fig.4}(c), we see that the energy is non-vanishing along the unstable manifold
of the (hyperbolic component of) the CS, and only the relative intensity is (slightly)
changed in time.  
These different intensities are enough to change the far-field emission as shown in
Fig.~\ref{fig.6}.  
An overall similar pattern is observed for both times, but with different intensities at
different emission angles. 
These results are in agreement with the interpretation at the end of Sec.~\ref{sec.nonH}
that the unstable manifolds of the hyperbolic and nonhyperbolic components of the CS are
close to each other  inside~$I$. 
However, the numerical simulations presented here suggests that differences in the
intensities inside~$I$ lead to far field emissions that are similar but not identical
for~$t<t_c$ and~$t>t_c$.

\section{Conclusions}\label{sec.conclusion}

This paper describes the chaotic ray dynamics in dielectric cavities in
terms of the theory of transient chaos.  
After properly taking into account the partial leakage introduced by Fresnel's law, the
long time escape and emission are found to be governed by a chaotic saddle
(CS)~\cite{gaspard,dorfman,ott,TG}. 
The energy inside the cavity is distributed according to the c measure~\cite{PY}, that is
non-zero along the unstable manifold of the CS~\cite{T}.  
In the generic case of cavities showing mixed phase space, it is useful to consider a
division of the CS in hyperbolic and nonhyperbolic components~\cite{JTZ,letter},
related to the exponential ($t<t_c$) and power-law~($t>t_c$) decays of energy inside the
cavity.  
For longer times ($t>t_c$) the energy concentrates in the nonhyperbolic component of the
CS, but the emission shows only small differences from shorter times~($t<t_c$).  
This is explained by the fact that, when the nonhyperbolic regions are away from the
critical line, the unstable manifolds of the hyperbolic and nonhyperbolic components are close to
each other inside the leak region.

From the point of view of previous investigations in dielectric cavities, these results
provides an unifying and more general description of experiments and simulations.  
For instance, the localization in the far field emission is shown here to be related to
the unstable manifold of the CS and not of a single periodic orbit as previously reported,
what is expected to be of practical relevance in cases where no simple periodic orbit exists
close to the critical line. 

From the point of view of dynamical systems theory, the novel aspects of this paper are
twofold: (I) it takes the partial leak of optical systems explicitly into account in the
definition (of the unstable manifold) of the CS (Sec. \ref{ssec.invariantsets}); and
(II) it shows how systems with mixed phase space can be effectively described
(Secs. \ref{sec.nonH} and \ref{sec.numerics}).
The success of point (II) in the description of physically relevant quantities, such as the
intensity of the far field emission, suggests
that future works in nonhyperbolic Hamiltonian systems should consider in more detail and
more rigorously the division of  the chaotic saddle in hyperbolic and nonhyperbolic
components~\cite{JTZ,letter}.  
Altogether, it is also interesting to see that many of the fundamental 
concepts of transient chaos theory
(e.g., the CS, its unstable manifolds, and c measure) achieve an experimental concreteness
in optical systems.  
This is perhaps only comparable to the case of fluid dynamics, where the unstable manifold
of the CS can be visually observed in scattering systems, specially when a constant
injection or activity of tracers compensate the natural decay through the fluid
flow~\cite{JTZ,schneider,report}. 

Similar to the case of fluids, experiments and applications in dielectric microcavities
are rarely a simple decay of rays that have been excited at some time~$t=0$, as considered
in the model of this paper.   
Instead, energy is constantly pumped to the system from outside through some gain in the
medium. 
The relevance of the rays that survive for long time inside the cavity by approaching the
CS is that their intensities are enhanced through the input of energy. Accordingly, in the wave picture,
the (high-Q) lasing modes concentrate in the region of the CS~\cite{casati}.
The details of these experimental mechanisms, and how they could be included in the ray picture, are
beyond the scope of this paper.  
However, generally one can think that these different mechanisms translate in a minimum
confinement time~$t_G$, and only rays confined for times $t>t_G$ are relevant.  
In this picture, the results of this paper suggest that depending whether $t_G < t_c$ or
$t_G > t_c$ the energy inside the chaotic component of the cavity changes dramatically: it
will be concentrated mainly in the hyperbolic ($t_G < t_c$) or nonhyperbolic ($t_G >
t_c$) component of the saddle (in optical microcavities $t_G\gg t_c$).  
However, the emission pattern is similar in these two cases because it is determined by
the unstable manifold of the CS inside~$I$.  
A similar emission pattern is also expected when the relevant modes are {\em inside} the KAM islands and
tunnel the (dynamical) barrier that separates them from the chaotic sea.

\acknowledgments
I thank M. Hentschel and T. T\'el for illuminating discussions.

\end{document}